\documentclass[12pt]{iopart}

\usepackage{graphicx}

\begin{document}

\title{Modeling capsid self-assembly: Design and analysis}

\author{D. C. Rapaport}

\address{Department of Physics, Bar-Ilan University, Ramat-Gan 52900, Israel}

\ead{rapaport@mail.biu.ac.il}

\date{June 16, 2010}

\begin{abstract}

A series of simulations aimed at elucidating the self-assembly dynamics of
spherical virus capsids is described. This little-understood phenomenon is a
fascinating example of the complex processes that occur in the simplest of
organisms. The fact that different viruses adopt similar structural forms is an
indication of a common underlying design, motivating the use of simplified,
low-resolution models in exploring the assembly process. Several versions of a
molecular dynamics approach are described. Polyhedral shells of different sizes
are involved, the assembly pathways are either irreversible or reversible, and
an explicit solvent is optionally included. Model design, simulation methodology
and analysis techniques are discussed. The analysis focuses on the growth
pathways and the nature of the intermediate states, properties that are hard to
access experimentally. Among the key observations are that efficient growth
proceeds by means of a cascade of highly reversible stages, and that while there
are a large variety of possible partial assemblies, only a relatively small
number of strongly bonded configurations are actually encountered.\\

\noindent{\it Keywords\/}: self-assembly, viral capsids, molecular dynamics
simulation

\end{abstract}

\pacs{81.16.Fg, 87.16.Ka, 02.70.Ns}

\section{Introduction}

The formation of the capsid shells that package
the genetic material of spherical viruses
\cite{cri56,cas62} is a particularly familiar instance of self-assembly in the
natural world, a phenomenon at the border between biology and
physics. The
rational design of antiviral agents would benefit from an improved understanding
of how such shells assemble, while in the industrial environment molecular scale
self-assembly is also expected to play a significant role in advancing
nanotechnology. More generally, for a broad range of applications, structure
formation in self-assembling molecular complexes is a particularly important
process \cite{whi02}.

Experimentally, direct observation of evolving molecular assemblies is
inherently difficult, with the final states revealing little about how they came
into being.
Strictly speaking,
self-assembly also implies a nonequilibrium state, thus predictive
theory (in the form of statistical mechanics) is absent. While simple mechanical
models have been used to explore steric aspects of capsid assembly \cite{cas80},
an approach based on the molecular dynamics (MD) simulation \cite{rap04bk} of
suitably designed models provides access to the assembly pathways themselves and
a means for examining the possible existence of universal organizational
principles that govern self-assembly. MD simulation is also able to predict the
time-dependent populations of partial assemblies providing, in principle, a
direct link with experiment \cite{end02}.

The highly symmetric capsid shapes are a consequence of their being assembled
from multiple copies of one or a small number of distinct capsomers
\cite{bak99}. Capsid assembly, a process whose details are little understood
\cite{fox94}, is governed by different classes of interactions: There are
interactions between the capsomers and the genetic material that initiate and
regulate assembly, and subsequently stabilize the end product. More relevant for
the present study are the protein-protein interactions between capsomers that
are also important in stabilizing the shell structure. What makes capsid
self-assembly suitable for simulation, despite the apparent biochemical
complexity, is the fact that it is able to occur reversibly {\em in vitro}
\cite{pre93,zlo99,cas04}, without the genetic material that is essential to the
virus {\em in vivo} (in other cases nucleic acid must be present \cite{spe95}).
In addition, structurally intact empty shells occur {\em in vitro} after removal
of their contents, and viruses themselves form empty capsids \cite{cas62}.
Modeling is simplified considerably by such information, since only a very small
number of well-characterized components need be considered.

Motivation for a reduced description stems from the robustness of self-assembly
\cite{cas80} and the prominence of icosahedral symmetry. Nature has adopted this
structural motif precisely because the high degree of symmetry leads to a
minimal set of construction specifications \cite{cri56}; in addition, the almost
spherical form offers near-maximal volume for a given surface area. Since the
task of the genetic information embodied in the viral nucleic acid is not only
to instruct the virus how to infect the host, but also to specify how it must
replicate itself, if less information can be devoted to the latter mission, more
will be available for the pernicious primary task. Analogous icosahedral motifs
are to be found in geodesic domes, whose detailed structures are a consequence
of the same minimalist construction specifications, as well as considerations of
optimal rigidity. Beyond their basic packaging role, capsomers also have an
important function in the virus life cycle \cite{joh96} that lies beyond the
scope of simple structural models.

MD modeling of capsid self-assembly, based on simple structural models that
retain sufficient detail to ensure meaningful behavior, was introduced in
Refs.~\cite{rap99,rap04a}. The principal characteristics of the approach are an
effective molecular shape formed out of rigidly arranged soft spheres that
enables particles to fit together in a closed shell, and multiple interaction
sites positioned to stabilize the correct final structure. This highly
simplified representation can be contrasted with real capsomers \cite{bak99}
that consist of intricately folded proteins whose exposed surfaces form
relatively complex landscapes. The initial focus of the simplified approach was
on achieving assembly; pathways were not investigated in detail, and solvent was
omitted on account of computational limitations. Shells of size 60, formed from
triangular and trapezoidal particles, were considered, the latter corresponding
to T=1 viruses; even larger shells of size 180, corresponding to T=3 viruses,
were also grown. While the later stages of the work involved reversible
assembly, a more reasonable approach from a physical perspective, the early part
was based on irreversible assembly, again in order to reduce the computational
effort. 

The next step, extending the simulations to include an explicit atomistic
solvent, had to await further increases in computing power \cite{rap08,rap10},
but even so the simulations were limited to icosahedral shells constructed from
triangular particles, rather than the larger shells considered previously.
Solvent presence aids cluster breakup without subassemblies needing to collide
directly, curtails the ballistic nature of the particle motion,
and serves as a heat bath to absorb energy
released by exothermal bond formation. The simulations revealed that
self-assembly proceeded by means of a cascade of reversible stages, with a
strong preference for low-energy intermediate states, eventually leading to a
high proportion of fully assembled shells. While it may seem paradoxical,
reversibility is the key to efficient production due to its ability to help
subassemblies avoid becoming trapped in states off the correct pathway. These
models, and some of their consequences, are explored below.

Other studies have also addressed capsid assembly dynamics. An alternative
particle-based MD simulation involved quasi-rigid bodies formed from hard
spheres \cite{ngu07}. Even simpler capsomer representations have been based on
spherical particles, rather than the extended particle shapes considered here,
with either directional interactions \cite{hag06} whose range exceeds the
particle size, or bonding energies determined by local neighborhood rules
\cite{sch98}; in these simulations the solvent is represented implicitly using
stochastic forces.
The motivation for using relatively complex particle designs, rather than simple
spherical particles, is the fact that capsomers themselves are extended bodies
whose customized shapes generally appear tailored to achieve properly assembled
shells. The practical benefit of using model particles whose size is comparable
to the interaction range is that the design ensures maximal bonding forces
across interfaces between correctly positioned and oriented particles, while
essentially eliminating bond formation in other situations. A similar outcome
using spherical particles with directional interactions is less readily
achieved, since the scope for tuning such designs to favor certain structural
motifs -- involving specific positions and orientations subject to narrow
tolerance ranges -- is more limited. Consequently, as demonstrated later, the
present results show a strong preference for correct assembly, and even on the
rare occasion that incorrect growth is initiated the unfavorable energetics
ensures the transience of such structural defects.

Monte Carlo simulations of patchy spheres \cite{wil07},
tapered cylinders \cite{che07}, and
particles with force centers located at the vertices of polygonal pyramids
\cite{joh10} represent further alternatives, but here the dynamical aspects of
assembly are absent. At the other extreme on the complexity scale are the
all-atom descriptions of capsomer proteins, although such MD simulations
\cite{fre06} extend over very short time intervals, sufficient for testing the
stability of preassembled shells. If the particle dynamics underlying
self-assembly is not a consideration, then a range of theoretical approaches for
studying capsid structure exist \cite{lid03,twa04,zan04,hic06,hem06}, and
experimental results have been interpreted using concentration kinetics
\cite{zlo99,van07}.

The present paper reviews some of the key aspects of the earlier MD work and
provides early results from an ongoing effort to extend the approach. The
organization is as follows: General considerations involved in modeling capsid
construction are introduced, with details of specific designs, based on rigid
soft-sphere assemblies, that give the particles their overall shape. The
interactions responsible for structure formation are described, including
alternative approaches based on both irreversible and reversible bonding.
Assembly scenarios are then covered, together with the interaction restrictions
that were necessary in the early stages of the work to supplement the
interactions in the case of irreversible bonding. Computational techniques are
summarized, followed by a selection of quantitative and visual results from past
and current simulations.

\section{Simulation methodology}

\subsection{Capsomer particle design}

There are few guidelines to aid in the design of model capsomer particles apart
from the need for the reduced representations to retain sufficient detail to
ensure meaningful behavior when used in MD simulation. The two principal
characteristics for the present work are the effective molecular shape, which
must be tailored to ensure particles can fit together to form closed polyhedral
shells, and the interactions between particles that drive self-assembly and
maintain the structural integrity of partial assemblies and complete shells. The
models introduced here are all based on rigid assemblies of soft-sphere atoms
that give the particles their shape, together with precisely positioned
attraction sites that ensure particle pairs are properly positioned and aligned
when in their minimal energy bound states. General thermodynamic considerations
aid in choosing the force parameters, but otherwise the approach is entirely
empirical and the parametrization is not tied to any particular experimental
measurements. Progress in discrete particle modeling of this kind is determined
by available computer power, with increasing computational capability allowing
the incorporation of additional features to enrich the model, as well as
permitting the larger systems and longer runs that may be needed to establish
behavior.

\begin{figure}
\begin{center}
\includegraphics[scale=0.20]{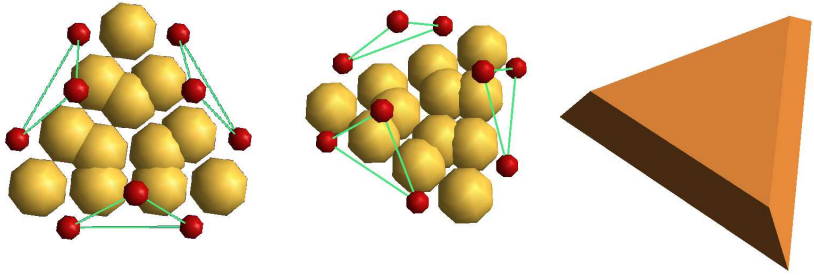}
\end{center}
\caption{\label{fig:01} (Color online) Details of the original model capsomer
particle, with the large spheres that provide the excluded volume and the small
spheres denoting attraction sites; the equivalent triangular block is also
shown, with beveled lateral faces that allow formation of a 60-faced polyhedron
from multiple copies of the particle, using irreversible bonding.}
\end{figure}

The initial study \cite{rap99} considered rigid triangular particles, shown in
Fig.\,\ref{fig:01}, designed for assembly of 60-faced pentakisdodecahedral
shells. The overall shape is produced by the arrangement of the larger spheres.
The repulsive force between spheres in different particles, used to prevent
spatial overlap, is based on the truncated Lennard-Jones potential
\begin{equation}
u(r) = \left\{
\begin{array}{ll}
4 \epsilon [(\sigma / r)^{12} - (\sigma / r)^{6} + 1/4] &
 \quad r < r_c = 2^{1/6} \sigma \\[4pt]
0 & \quad r \ge r_c
\end{array}
\right.
\end{equation}
where $r$ is the sphere separation, $\sigma$ approximates the sphere diameter,
$\epsilon$ determines the energy scale, and $r_c$ is the interaction cutoff. In
the reduced MD units used here, $\sigma = \epsilon = 1$; thus the spheres that
give the particles their shape are of approximately unit diameter. While there
are multiple interactions between the spheres belonging to particles that are in
close proximity, the computations based on pair potentials are simpler than
evaluating the overlap of complex rigid bodies required by alternative shape
representations.

The small spheres in Fig.\,\ref{fig:01} signify attraction sites, three per face,
and extending beyond the volume occupied by the spheres that give the particle
its shape (towards the shell interior when particles are correctly positioned);
corresponding pairs of attraction sites on different particles interact through
an inverse-power potential and at close range this interaction smoothly merges
into a narrow harmonic well (see details below). Successful assembly involves
the coupling of each of the complementary pairs of attraction sites, and is
conditional upon particles having the appropriate dimensions to ensure
components fit together, with the angles of the planes containing the attraction
sites determined by the overall shell shape.

\begin{figure}
\begin{center}
\includegraphics[scale=0.20]{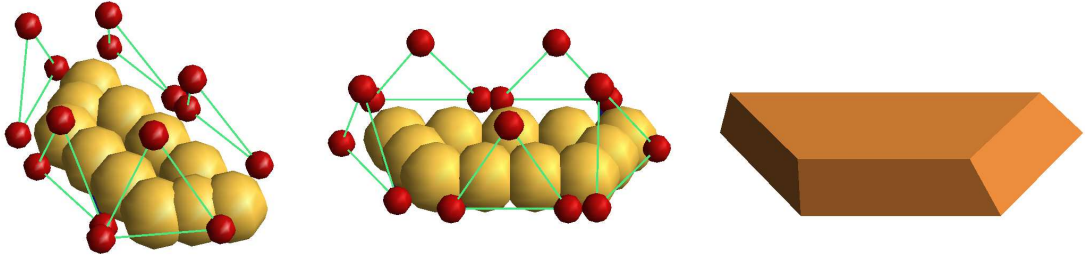}
\end{center}
\caption{\label{fig:02} (Color online) Capsomer model used for T=1 assembly with
irreversible bonding; the spheres comprising the particle and the effective
trapezoidal shape are shown.}
\end{figure}

Subsequent work \cite{rap04a} involved models that extended this basic design.
Shells corresponding to a T=1 virus are constructed from 60 identical
trapezoidal capsomer particles shown in Fig.\,\ref{fig:02}. The shell can be
regarded as an icosahedron \cite{wil79} each of whose 20 equilateral triangular
faces is subdivided into three coplanar trapezoidal units. The lateral particle
faces within the triangle are normal to the triangular plane, whereas those
along the outside of the triangle are inclined at $20.905^\circ$ to the normal,
resulting in a dihedral angle of $138.190^\circ$; these angles dictate the
positioning of the attraction sites. Each of the three short edges contains a
single set of attraction sites, while the long edge contains two sets. 

\begin{figure}
\begin{center}
\includegraphics[scale=0.20]{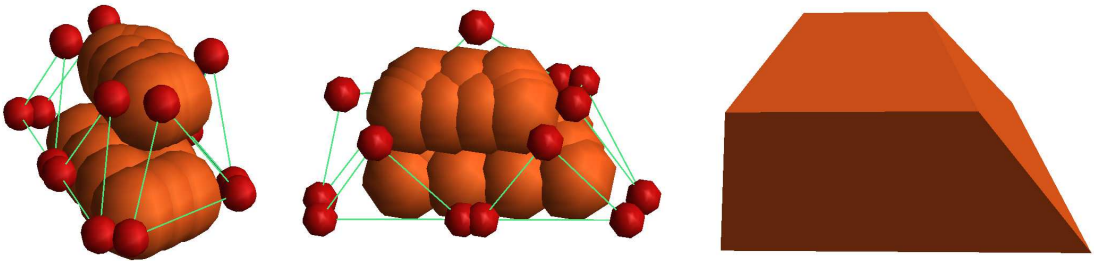}
\end{center}
\caption{\label{fig:03} (Color online) Views of T=3 capsomer model.}
\end{figure}

A further increase in complexity is the T=3 shell consisting of 180 trapezoidal
particles. This shell is based on a rhombic triacontahedron \cite{wil79} with 30
identical rhombic faces; each face is subdivided into two isosceles triangles
(the base angles are $58.283^\circ$, so the triangles are almost equilateral),
and each of these triangles is then divided into three coplanar trapezoidal
particles. The lateral faces within the same triangle, and between the triangles
comprising the rhombus, are normal to the triangular plane, while the other
faces are inclined at $18^\circ$, producing a dihedral angle of $144^\circ$. The
particle is shown in Fig.\,\ref{fig:03}. Note that two layers of spheres are now
used to increase the overall thickness of the particle in order to help avoid
unwanted interactions.

Since 60 is the maximum size under conditions of complete equivalence, the
construction of larger structures is explained by employing the concept of
quasi-equivalence \cite{cas62,cas85,ros85}. Quasi-equivalence requires an
autosteric mechanism, in which capsomer protein conformation varies by a small
amount \cite{red98} depending on the position in the shell (mechanical analogies
are discussed in Ref.\,\cite{cas80}). However, the model designs used here assume
a fixed shape. To overcome this difficulty, three particle variants with
slightly different face angles are used (with attractive interactions only
between corresponding face pairs), each destined to occupy a different subset of
locations (corresponding to the subdivision of the isosceles triangles into
trapezoids) within the shell. 

\begin{figure}
\begin{center}
\includegraphics[scale=0.25]{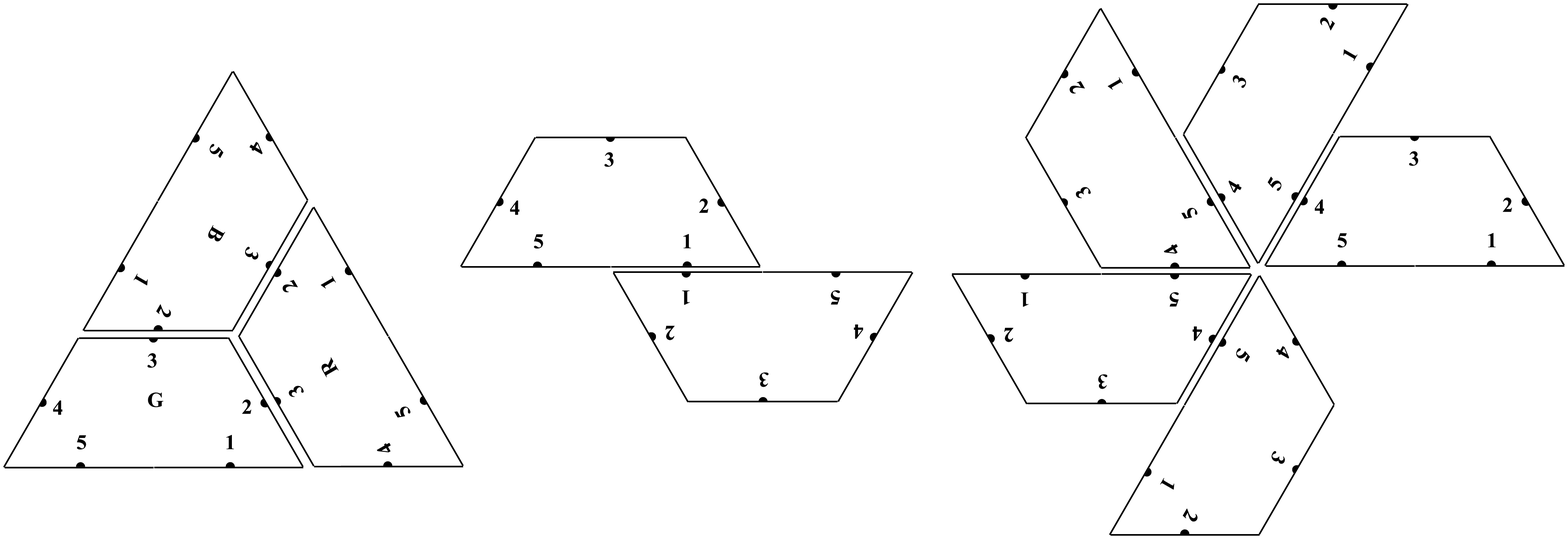}
\end{center}
\caption{\label{fig:04} Trimer, dimer and (opened-up) pentamer configurations
with labeled attraction sites; the color identification (used for T=3) is
included in the triangular configuration.}
\end{figure}

The labeling scheme used for the sets of attraction sites is shown in
Fig.\,\ref{fig:04}. For T=1 the trimer takes the form of an equilateral triangle,
whereas for T=3 a small change of apex angle is needed that makes the triangle
isosceles. Since color coding is useful for T=3, particles are labeled B, G, R
(blue, green, red) following \cite{spe95}. The five sets of attraction sites
follow the same bonding pattern in each particle, namely, complementary site
pairs in sets 2 and 3 can bond, as is also the case for sets 4 and 5, while
sites in sets 1 bond with each other. Different site pairings are associated
with trimer, dimer, and pentamer or hexamer formation. Three particles joined by
2-3 bonds produce a planar triangular trimer. The 1-1 bonds form a dimer; it is
nonplanar for T=1, while for T=3 it is planar when two type G particles are
involved and nonplanar in the alternative R-B case. The 4-5 bonds produce a
nonplanar, flower-like pentamer for T=1. On the other hand, for T=3, there are
again two possibilities; if all particles are of type B they produce the
pentamer, but a hexamer is formed when alternating R and G types are involved
(appearing as three coplanar pairs).

\begin{figure}
\begin{center}
\includegraphics[scale=0.20]{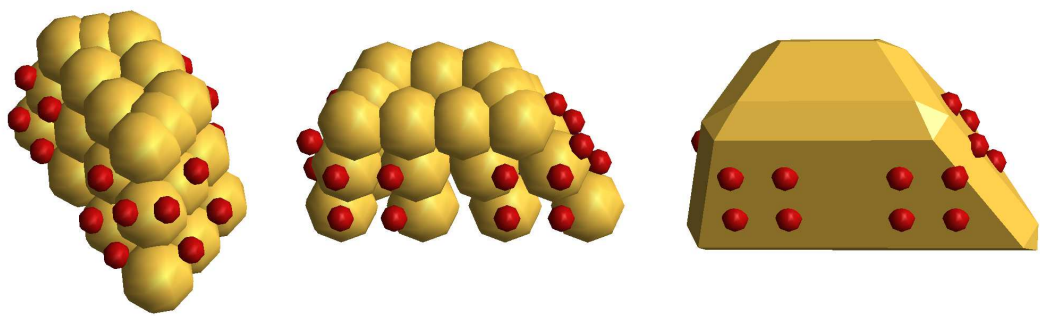}
\end{center}
\caption{\label{fig:05} (Color online) Views of capsomer model used for T=1
reversible bonding.}
\end{figure}

The particles described so far were used in studies of irreversible assembly.
Reversible bonding was also considered in Ref.\,\cite{rap04a}.
Fig.\,\ref{fig:05} shows the design used for T=1 shells in this case.
Constructing the particle from three layers of spheres reduces even further the
likelihood of incorrect bonding; this is now a more important issue since there
are no restrictions (see below) to help avoid attractive interactions that do
not contribute to the final shell. Each bond now involves four attraction-site
pairs; the energetic gain of a correctly aligned state is enhanced by
distributing the interactions over more site pairs. Owing to the increased
particle thickness, attraction sites can now be positioned so they no longer
extend beyond the actual area of the lateral faces; the additional steric
screening helps prevent unwanted interactions. Such benefits come at the price
of increased computational effort.

\begin{figure}
\begin{center}
\includegraphics[scale=0.7]{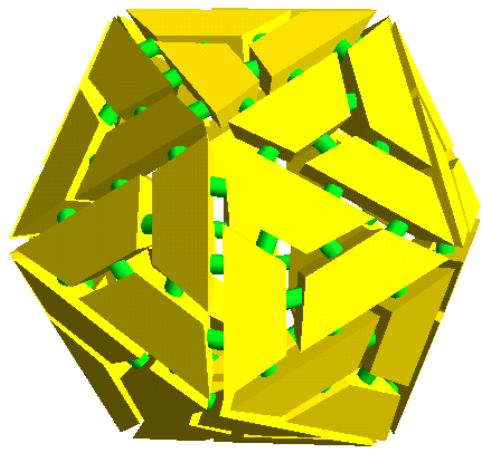}
\end{center}
\caption{\label{fig:06} (Color online) Complete T=1 shell with 60 particles;
particles are shown slightly reduced in size to allow the bonds to be seen.}
\end{figure}

\begin{figure}
\begin{center}
\includegraphics[scale=0.7]{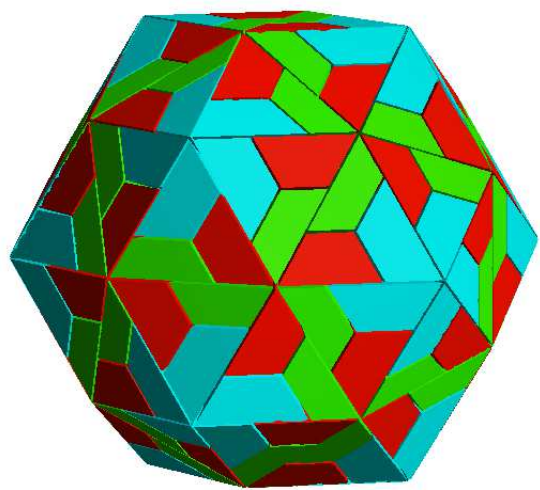}
\end{center}
\caption{\label{fig:07} (Color online) Complete T=3 shell with 180 particles of
three (color-coded) types.}
\end{figure}

Figs.~\ref{fig:06} and \ref{fig:07} show the complete shells that these
particles are designed to form (the images are of shells produced by the
simulations). The former shows a T=1 shell; due to the nature of the bonding
forces (see below) complementary attraction sites coincide in the ground state,
although in the figure particle size is slightly reduced so the bonds can be
seen. The latter shows a T=3 shell formed from the three different particle
types; here particles are drawn to show their effective sizes, hiding the bonds.
The different local particle arrangements described previously are all visible.

\begin{figure}
\begin{center}
\includegraphics[scale=0.25]{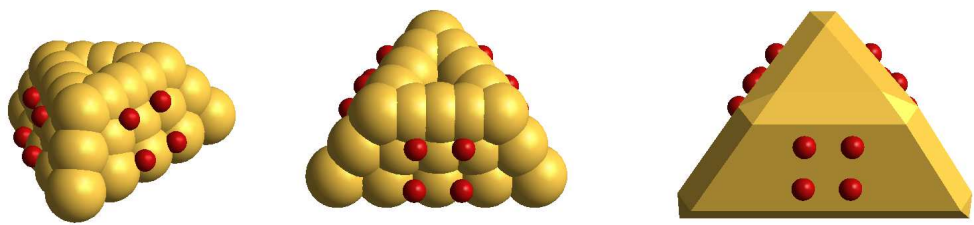}
\end{center}
\caption{\label{fig:08} (Color online) Model particle and its effective
truncated pyramidal shape used for assembly with an explicit solvent.}
\end{figure}

More recent simulations \cite{rap08,rap10} involving both reversible bonding and
an explicit solvent have been confined to icosahedral shells, the reason once
again being computational. Particles have an effective shape of a truncated
triangular pyramid, shown in Fig.\,\ref{fig:08}, with multiple layers of spheres
again used for steric shielding of the attractive interactions. The lateral
faces containing the attraction sites are inclined at $20.905^\circ$. The volume
of this particle should be contrasted with the one shown in Fig.\,\ref{fig:01}
that consists of a single layer of spheres.

These represent the simplest of designs and there are, of course, numerous ways
of enhancing the models. One example is the introduction of more complex
particle surfaces, allowing the use of a `lock and key' mechanism to represent
additional steric effects. An attempt was made to use such a technique in early
work on small triangular particles, by adding an extra sphere to the center of a
lateral face and leaving a vacancy in the complementary face, but this mechanism
failed to provide sufficient extra rigidity. The subsequent increased size and
multiple attraction sites of the larger particles appear adequate for the
present work.

\subsection{Bonding interactions}

Bond formation between real capsomers involves contributions from a relatively
large number of pairs of interaction sites distributed over the mutual contact
surface between the proteins. Although such interactions are not individually
directional, in view of the substantial size of the capsomer compared to the
effective interaction range, it is unlikely (and disadvantageous) that the
interactions would be able to produce strongly bound states in which capsomers
are incorrectly aligned. This is not necessarily true for simplified models with
few attraction sites and an overall capsomer particle size similar to the
interaction range; in this case, small clusters can become trapped in states
corresponding to spurious local energy minima instead of developing into
complete shells. Ensuring defect-free assembly can be accomplished by a trade
off between model particle size and interaction complexity (with possible
supplementary restrictions governing when and where attraction is allowed). The
models used in the present series of studies demonstrate how increasing the
particle size (at increased computational cost) relative to the interaction
range allows the interaction details to be simplified (and other artifacts
eliminated).

Two fully-bonded particles are held together by interactions between multiple
(three or four) pairs of complementary attraction sites. Typically, particles
might be drawn together initially by just one of the attraction site pairs, and
they then reorient so the remaining site pairs can participate. The use of
multiple sites, where only complementary sites can interact, helps accomplish
several goals: (a) The orientation of a lateral face and, consequently, the
preferred dihedral angle between particles, is determined by the plane
containing the sites. (b) The overall bonding interaction is distributed over
the contact face; this ensures that the total bonding energy of misaligned
particles can only be a fraction of the value in the ground state, thereby
reducing the stability of improper bonds. (c) Multiple attraction sites enhance
structural rigidity by suppressing internal degrees of freedom such as twisting
or flapping.

The functional form of the interaction between bond-forming attraction sites,
chosen entirely for convenience, is a negative power of the site separation $r$
(e.g., $1 / r^2$) when the sites are not too close, and for $r < r_h$ it takes
the form of a narrow harmonic well; in the minimum energy state $r = 0$ and the
attraction sites coincide. In the case of reversible bonding, the force is
derived from the potential
\begin{equation}
u(r) = \left\{
\begin{array}{ll}
e (1 / r_a^2 + r^2 / r_h^4 - 2 / r_h^2) & \quad r < r_h \\[4pt]
e (1 / r_a^2 - 1 / r^2) & \quad r_h \le r < r_a
\end{array}
\right.
\end{equation}
with typical parameter values $e = 0.1$ (in the case of reversible bonding the
effect of varying the interaction strength parameter $e$ is investigated -- see
below), $r_h = 0.3$, and cutoff $r_a = 3$ (in the solvent-free work $r_a = 2$).
Particle size exceeds the interaction range, although less so than in real
capsomers; the effect of a relatively short-ranged force, together with multiple
attraction sites, is to reduce the attraction between wrongly positioned or
oriented particles.

The interactions used with irreversible bonding are similar, the principal
difference being that a permanent bond forms the first time $r < r_h$. Following
the formation of this bond only the harmonic component of the attraction acts,
irrespective of $r$, implying an infinitely high barrier to bond breakage.
Earlier work also included an explicit torsional interaction to accelerate the
bonding process, but this was eliminated once it became apparent that multiple
pair interactions alone were adequate for achieving correct particle
orientation.

Preferred assembly pathways are another feature meriting study. Since the
likelihood of two arbitrary subassemblies being able to mesh successfully is low
(unless incompatible pieces can be discarded in the process -- possible only if
bonds can break without too much effort), a hypothetical assembly scenario can
be based on a multistage process, with small clusters of particles having a
specified organization forming initially, and then combining into increasingly
larger subassemblies. These small clusters must be able to `tile' the complete
shell, so a possible first stage is the assembly of triangles from the
trapezoidal particles, while in the second stage these trimers bond to form full
shells. Experimental signatures of multistage assembly would be certain
preferred intermediate cluster sizes, or the more accessible rate concentration
dependence \cite{zlo99,zlo00}. While it is not possible to enforce such behavior
strictly by choice of interactions alone (it can be achieved using assembly
restrictions -- see below), increasing the interaction strength $e$ for those
bonds responsible for (e.g.) trimer development can help bias the growth pathway
in the correct direction.

\subsection{Computational details}

General MD methodology is discussed in Ref.\,\cite{rap04bk}; a brief summary of
the issues relevant to capsomer self-assembly simulation follows. Interaction
calculations are carried out using neighbor lists, with the list construction
following the procedure used for monatomic fluids. Separate lists are used for
the short-range repulsive forces between the spheres that give the particles
their shape, and for the longer-range forces between attraction sites. The
rotational equations of motion employ standard rigid-body methods; these,
together with the translational equations, are solved using a leapfrog
integrator, with a time step of $5 \times 10^{-3}$ (in reduced MD units).

The earlier simulations did not include a solvent (a simplification motivated by
the fact that it is often adopted in modeling protein folding); even the
simplest atomistic solvent would increase the computational effort
substantially, not only because of the additional elements involved in the
simulation, but also because of the slower particle movement when in a solvent
rather than in a vacuum. To ensure an adequate supply of unbonded particles in
the solvent-free simulations, partial shells below a certain size threshold were
broken up at regular intervals (with a frequency depending on growth rate) by
switching off their bonding interactions for a short period.

Later simulations included an explicitly modeled solvent. The solvent atoms are
identical to the particle spheres, and are subject to the same soft-sphere
repulsion. The mass of each particle is proportional to its volume, here
typically $21\times$ the solvent atom (whose mass is unity in MD units); having
a much smaller mass ratio than in real viruses reduces the assembly timescale,
making it more accessible to MD without any qualitative change in behavior
expected. Particle concentration is much higher than in reality, but there is
adequate solvent present to ensure that the ballistic particle motion of the
solvent-free case is now dominated by diffusion.

Although the interactions between particles incorporate an attractive component,
the only interactions involving solvent atoms (both among themselves and with
the particles) are due to excluded volume; while this is a major simplification
compared to actual capsomers in aqueous solution, it is reasonable to expect
that the essential features of self-assembly are preserved. A more focused
program of experiment and simulation will be needed for systematically
determining which properties of the capsomers and their environment are
necessary for the assembly process and which play a more passive role.

Two kinds of boundary conditions have been used. Because of the importance of
visualization (including animation), early work used elastically reflecting hard
walls, implemented using short-range repulsive forces that act perpendicularly
to the surfaces; this avoids the sometimes confusing imagery that accompanies
periodic boundaries. Later work, with larger systems, returned to conventional
periodic boundaries, with the minor visual artifacts associated with particles
residing close to region boundaries no longer a significant issue.

In the initial state, particles and solvent atoms (where present) are positioned
on a lattice; each is assigned a random velocity, and the particle orientation
is also set randomly. The lattice spacing is determined by the overall number
density of the system; if there is a risk of overlap at the start, particles can
begin as collapsed objects (the size of a single sphere) and allowed to expand
gradually to their final shape over the initial (e.g.) 5000 steps.

Exothermal bond formation will heat the system; this problem is particularly
acute when there is no solvent due to the limited number of degrees of freedom
capable of absorbing the excess thermal energy. Applying a weak damping force,
$- \gamma \,(\vec{v} \cdot \vec{r})\, \vec{r} / r^2 $, along each bond resolves
this issue, where $\vec{v}$ is the relative velocity of the attraction sites and
the damping coefficient $\gamma = 0.1$ (typically). Use of constant-temperature
MD then ensures that the overall temperature does not change despite bond
formation and damping; the net effect is to transfer energy associated with
internal cluster vibration to the motion of entire clusters. The problem is less
severe when a solvent is included, but the thermostat is still required, and is
used to maintain a temperature of 0.667, corresponding to unit mean particle
kinetic energy (translational and rotational); the damping force is no longer
required in this case.

The interaction parameters are chosen to achieve efficient self-assembly while
maintaining numerical stability; there is presently no relation to experimental
association energies \cite{red98}. The number density $\rho$ affects the outcome
and must also be established empirically (here $\rho \le 0.2$). Too high a 
value of $\rho$ will not provide adequate space for shells to grow without
mutual interference, whereas if $\rho$ is too low, growth is retarded due to
particles tending to lie beyond their attraction range and the lack of
collisions that can, in the case of reversible bonding, help break off
incorrectly bound pieces from partially formed shells.

\subsection{Supplementary interaction restrictions}

In the initial work, confined as it was to smaller particles, with severe
computational limitations demanding rapid assembly and high yield (not
necessarily a requirement {\em in vivo}), the need to resort to irreversible
bonding for efficiency reasons required supplementary restrictions as to when
bonding interactions were allowed to act. Their purpose is the avoidance of
construction errors that could not be rectified subsequently; these errors, in
turn, would lead to mutant structures and even amorphous clusters.

With the benefit of considerably more powerful computers (not to mention a
little hindsight), these additional conditions are no longer needed for modeling
assembly at the current level of description, with bonds now simply regarded as
potential wells of finite depth. Nevertheless, it is worth considering these
restrictions; the ideas they embody may prove useful at some future stage, and
some of the mechanisms themselves may be physically justifiable as they could
manifest themselves through local conformational variation in response to
changes in bonding state. The alternative approaches based on irreversible and
reversible bonding represent, respectively, the extremes of kinetically limited
and equilibrium assembly. 

One simple restriction is that after a permanent bond has formed, the attraction
sites involved no longer interact with other particles. Since several site pairs
are involved in drawing a pair of particles into a bound state and a certain
amount of time can elapse between the first and last pairings, the risk of
forming incompatible bonds is reduced if neither of these particles can attract
new particles until bonding of all the site pairs is complete. If construction
follows a pathway in which, for example, dimers form initially, and then bond
into larger structures, an analogous condition is applied to entire dimers.

To minimize any adverse effects, if bonding fails to complete within a
prescribed interval (here, 4000 time steps) the existing partial bond is broken;
immediate rebonding of the separated particles is prevented by requiring them to
wait 1000 steps before becoming available for bonding again. The aim of
additional mechanisms of this kind is to help ensure the release of particles
that are unable to bond completely, as in the case when two particles attempt to
bond along different exposed faces of a cluster opening big enough for just one
of them. Yet another restriction, introduced for computational convenience, is
that particles permanently bound into different partially formed structures do
not attract one another, but only unbound particles; this avoids issues of
structures not being able to mesh correctly. Finally, to ensure that enough
particles are available for populating the complete shells, only a limited
number of subassemblies are allowed to begin the growth process.

Other interaction restrictions can be used to enforce assembly pathways; thus if
growth occurs via trimer intermediates, particles are first required to form
trimers, and only clusters with all their internal bonds complete are allowed to
associate into larger structures. Additionally, by restricting the number of
larger subassemblies that can nucleate (e.g.) by the joining of two trimers,
equivalent to a rate-limiting process \cite{zlo99}, it is possible to enhance
the yield of complete shells rather than numerous small fragments. Owing to the
intrinsic bond flexibility, the need for further restrictions only becomes
apparent as incorrect structures appear in the simulations; in some cases the
restrictions are used to enforce particular bonding sequences to prevent
partially bonded particles from encountering inappropriate bonding partners
\cite{rap04a}. As indicated earlier, use of reversible bonding eliminates such
issues.

\section{Results}

\subsection{General aspects}

System sizes and run lengths varied over a wide range. In the early work, with
irreversible bonding and no solvent, the T=1 simulations used 1000 particles,
with 13 shells allowed to develop and run lengths of up to $3 \times 10^5$ time
steps, while for the larger T=3 shells there were 4096 particles, 10 shells
allowed to develop and run lengths up to $8 \times 10^5$ steps. The T=1 case,
with reversible bonding and no solvent, also included 4096 particles, but the
run length was extended to $10^7$ steps; while there was no limit on the number
of shells allowed to grow, partial assemblies of size $\le 30$ were broken up
every $5 \times 10^5$ steps by turning off their attractive interactions over
the subsequent $10^4$ steps. At the other extreme, simulations involving
reversible bonds and a solvent consisted of 1875 triangular particles --
sufficient for 93 full shells -- and solvent atoms bringing the total to
125\,000; the required run lengths sometimes exceeded $70 \times 10^6$ steps.
Thus, between the original and the latest work, the overall computational effort
per run expanded over several orders of magnitude, with the longest runs
requiring extended periods (measured in weeks) of computation using workstations
with dual 3.6GHz Intel processors able to compute approximately $10^6$ steps per
day.

The selection of results shown here for the early solvent-free work is limited
to images produced during the simulations that lead to the growth of T=1 and T=3
shells. The main focus of more recent simulations that include solvent is on the
growth of smaller shells; the analysis carried out for these systems is more
detailed, with emphasis on quantitative results. Current simulations, described
briefly at the end of this section, attempt to address larger shells once again.

To allow post-analysis of the simulations, a sequence of `snapshots' is recorded
at intervals of 2000 steps over the duration of each run. Shell properties can
be measured directly from this retained data, allowing the average growth
statistics to be analyzed, together with the behavior of individual shells. Time
resolution is limited by the snapshot interval, so that shortlived bonds between
snapshots will be overlooked, as will merged events such as two added monomers
that are indistinguishable from dimer addition. However, since particles move
relatively slowly on the MD time scale, significant changes generally take
sufficiently long for the majority of individual growth steps to be identified.

\subsection{Shell analysis}

Algorithmically establishing the identities of particles that are members of
partial assemblies and verifying that shells are correctly assembled requires
the capability for identifying bound clusters \cite{rap04bk} and checking the
connectivity of their bond networks. With reversible bonding, there is a certain
amount of arbitrariness in defining what constitutes a bound particle pair.
Individual attraction sites are regarded as bonded when they are separated by a
distance $< 0.6$ ($= 2 r_h$), an empirically chosen threshold that avoids
transient bond breakage and reformation due to the thermal vibrations. Particles
are considered bound if all four complementary site pairs are bonded, a state
that also implies alignment owing to the relatively tight tolerances in the
design. In view of the comparative rigidity of multiply-bonded subassemblies, a
cluster corresponds to a closed shell if the numbers of particles and bonds
equal the expected values, a criterion that can also be verified visually.
Completely closed shells are subject to only the smallest of fluctuations in the
relative particle positions and orientations.

Describing the nature of incomplete shells, while straightforward when the
imagery is available -- a nearly complete shell is readily characterized, as is
a shell with a localized defect -- is not obviously quantifiable. Since
partially formed structures, and defect-free shell fragments in particular, have
a variety of morphologies, automated classification is a nontrivial task. Each
such structure can be represented as a bonded network, or graph, and while it is
possible to determine the graph topology by examining connectivity, and the
cluster compactness by counting missing bonds, it is not apparent how such
information can be utilized. Furthermore, it turns out that growth is not
necessarily a process in which shells grow monotonically by accretion of
individual particles or small subassemblies; indeed, larger subassemblies can be
seen to aggregate and (under reversible bonding conditions) groups forming
partial shells do manage to break away from larger structures. Results obtained
from the limited quantitative techniques available are described here, and these
will be followed by visual examples of some of the more interesting, and
complex, assembly events.

The most important observation regarding the overall behavior, applicable to
both irreversible and reversible bonding alike, is that polyhedral shells have
little difficulty growing to completion, and mutant structures are highly
unlikely. Partial shells tend to have few voids in their surfaces, and shells
nearing completion typically have a minimal number of openings; more open
cagelike structures with multiple lacunae tend to be avoided. The results also
reveal, not surprisingly, that an inappropriate choice of interactions, or even
a slight error in defining particle geometry, leads to a wide variety of
alternative structures, including open networks, incorrectly linked assemblies
of shell fragments and other ill-characterized shapes.

A goal of this work is to identify particle designs that produce a high yield of
properly assembled shells. The fact that incorrect assembly features more
prominently in \cite{hag06} is attributable to the fact that spherical particles
with directional interactions are limited in their ability to exclude
incorrectly bonded structures, while in \cite{ngu07} a similar outcome is due to
the inherent flexibility of the linked structures that are used instead of rigid
particles. Even further reduction of the structural preferences, as in the case
of tapered cylinders \cite{rap04a}, leads to a broad distribution of shell
sizes, since packing considerations are only able to impose partial ordering on
the particles, the exact opposite of the strong structuring of the present
approach.

\begin{figure*}
\begin{center}
\includegraphics[scale=0.75]{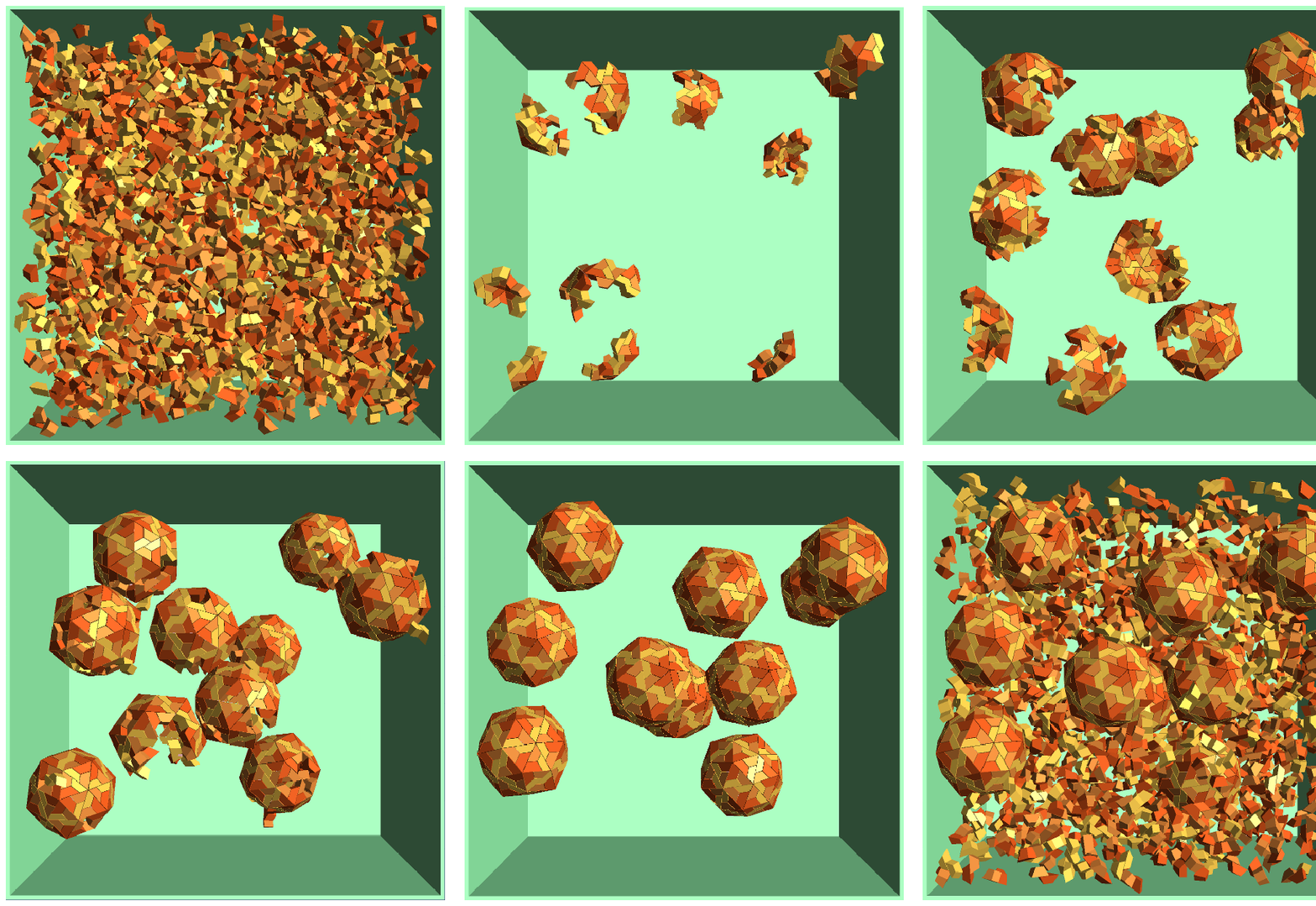}
\end{center}
\caption{\label{fig:09} (Color online) Snapshots from a T=3 simulation with
irreversible bonding: the frames show an early state, four views at different
times including the final state showing only the shells, and the entire system
in the final state (particles are color-coded by type).}
\end{figure*}

\begin{figure*}
\begin{center}
\includegraphics[scale=0.75]{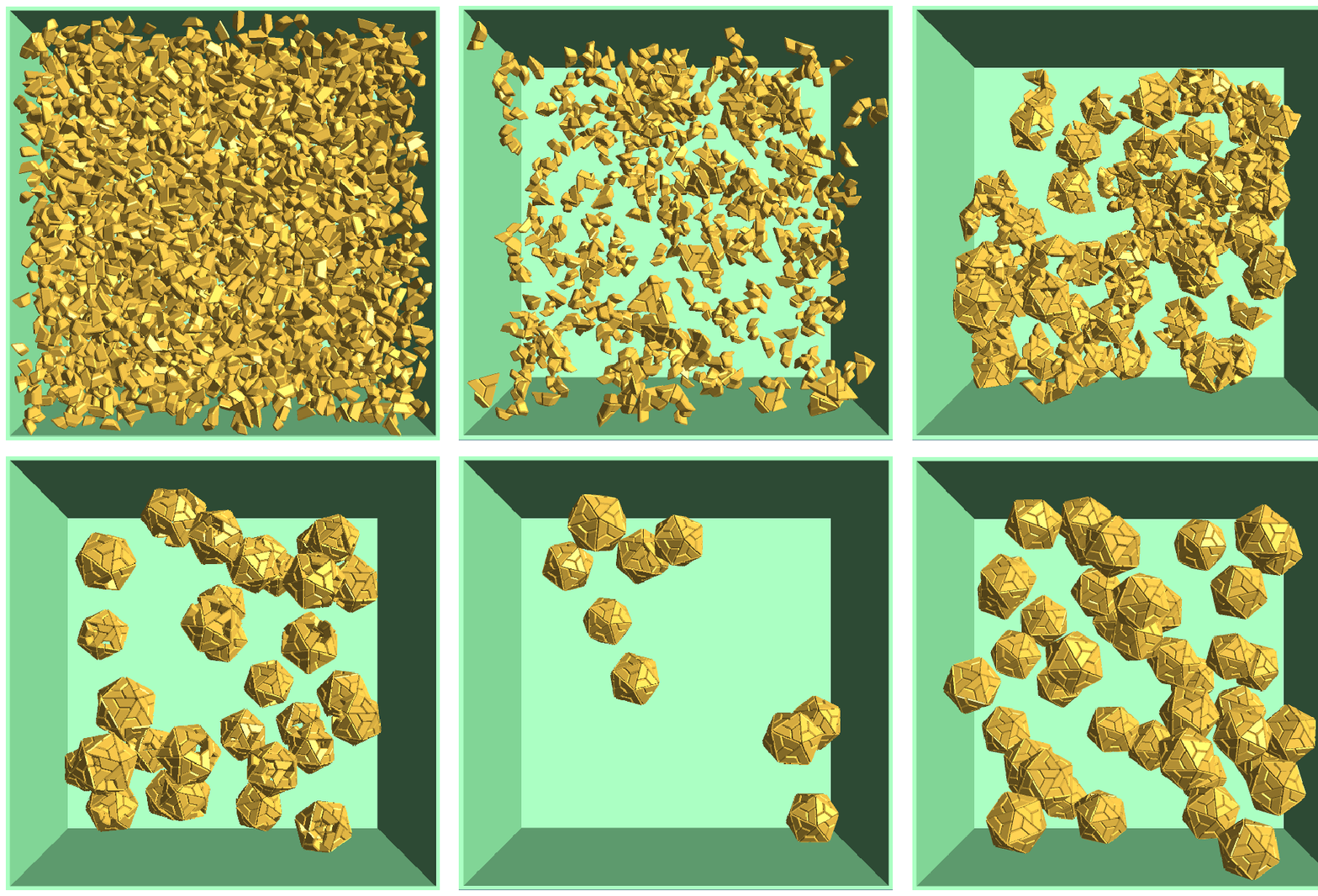}
\end{center}
\caption{\label{fig:10} (Color online) Snapshots from a T=1 simulation with
reversible bonding but without solvent: the frames show the entire system near
the start of the run, 366 clusters of size $\ge 2$ after $10^5$ steps (monomers
are not shown), 68 clusters of size $\ge 20$ after $3 \times 10^5$ steps
(smaller clusters are not shown), 33 clusters of size $\ge 50$ after $10^6$
steps, 9 complete clusters (size 60) after $3 \times 10^6$ steps, and final
state with 46 complete clusters after $10^7$ steps.}
\end{figure*}

The stored snapshot data can be used to produce animated sequences that provide
condensed summaries of the runs in full three-dimensional detail; this is very
helpful for extracting more subtle, and not readily quantifiable features of the
growth histories, examples of which are discussed below. Fig.\,\ref{fig:09}
shows several images of T=3 growth with irreversible bonding and a
dimer-weighted pathway, as extracted from the snapshot sequence (the color
scheme differs from Fig.\,\ref{fig:07}). Images from the much longer,
reversible, solvent-free T=1 case are shown in Fig.\,\ref{fig:10}.

\subsection{Shell growth}

The viability of the method is based on its ability to produce complete shells.
Examples of the successful outcome of solvent-free simulations, both
irreversible and reversible, have been described visually. The more detailed
quantitative results that follow are for icosahedral shell growth under
reversible bonding conditions and in the presence of a solvent. In this series
of simulation runs the dependence on the interaction strength $e$ is studied
systematically. If the range of variation is not too large this is equivalent to
examining the temperature dependence of the behavior. While the coverage of the
phase diagram is limited (other parameters defining the model have been sampled
more sparsely), it is more than adequate for demonstrating a selection of very
different outcomes.

Table~\ref{tab:1} provides a concise summary of these runs. The results are
expressed as the mass fraction of particles contained in complete shells, in
clusters of different size ranges, and the remaining monomers. Runs are
sufficiently long for the cluster populations to stabilize.

\begin{table}
\caption{\label{tab:1} Final cluster distributions for different interaction
strengths $e$; the results show the monomer mass fraction, the clusters grouped
into two size ranges, and the complete shells, with the maximum mass
fraction for each run in bold.}
\begin{center}
\begin{tabular}{lcccc}
\hline
       &       \multicolumn{4}{c}{Cluster mass fraction}               \\
 $e$   &Size: 1           & 2-14   &         15-19   &	       20      \\
\hline
 0.11  &  \textbf{0.7931} & 0.1157 &         0.0165  &         0.0747  \\
 0.115 &  \textbf{0.5153} & 0.1013 &         0.0101  &         0.3733  \\
 0.12  &  	  0.3040  & 0.0346 &	     0.0000  & \textbf{0.6614} \\
 0.125 &  	  0.1915  & 0.0518 &	     0.0101  & \textbf{0.7466} \\
 0.13  &  	  0.0709  & 0.0278 &	     0.0160  & \textbf{0.8853} \\
 0.14  &  	  0.0011  & 0.0774 &	     0.2922  & \textbf{0.6293} \\
 0.15  &  	  0.0000  & 0.1990 & \textbf{0.5983} &         0.2027  \\
\hline
\end{tabular}
\end{center}
\end{table}

At a low value, $e = 0.11$, there is practically no growth, due to minimal
initiation. The yield increases with $e$, and production efficiency peaks at
$e=0.13$, with a yield of 83 shells, out of a maximum possible 93. At higher $e$
-- 0.14, and especially 0.15 -- the ability to reach completion is inhibited by
excessive early growth, resulting in too many monomers being incorporated into
clusters prematurely. No oversized (mutant) clusters are encountered here,
although they would be expected for sufficiently large $e$. While similar
overall behavior is seen in reaction kinetics studies \cite{zlo99}, provided
nucleation is rate limited, MD needs no restrictions of this kind.

\begin{figure*}
\begin{center}
\includegraphics[scale=0.95]{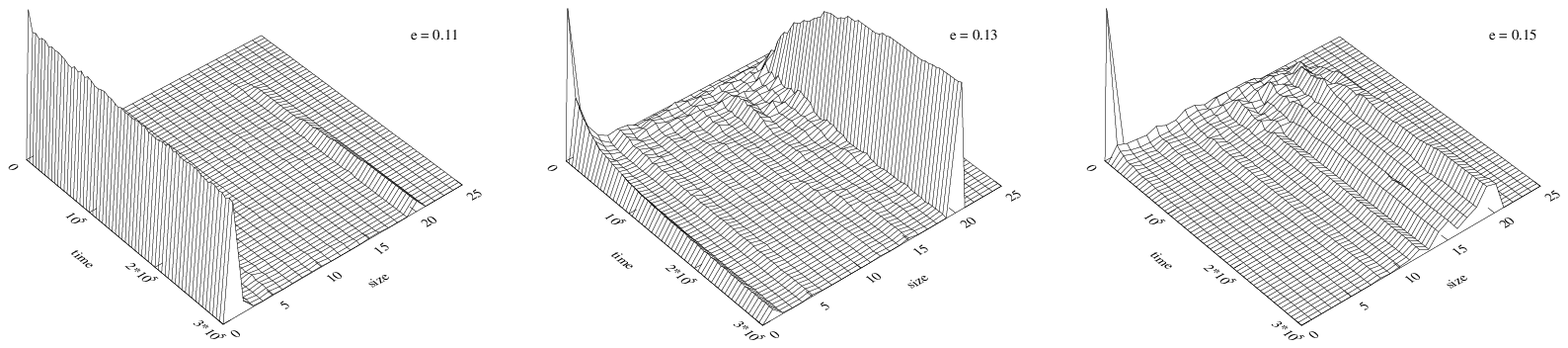}
\end{center}
\caption{\label{fig:11} Cluster size distributions as functions of time (MD
units); the distributions, including monomers, are expressed as mass fractions,
and $e$ (attraction strength) values are selected to show three distinct growth
scenarios.}
\end{figure*}

Examples of the time-dependent cluster size distributions appear in
Fig.\,\ref{fig:11}. The values of $e$ shown are selected to illustrate the
different kinds of behavior. When closed shells are produced in significant
number, the yield curves (i.e., the mass-fraction values for size 20) as a
function of time have the familiar sigmoidal shape -- rapid production after an
initial delay that eventually tapers off. It is also abundantly clear from
Table~\ref{tab:1} and Fig.\,\ref{fig:11} (see also \cite{rap08}) that in high
yield runs there
are
very few clusters of intermediate size remaining at the end,
essentially only complete shells and monomers. This is one of the key
observations to emerge from these simulations.

\begin{figure}
\begin{center}
\includegraphics[scale=0.20]{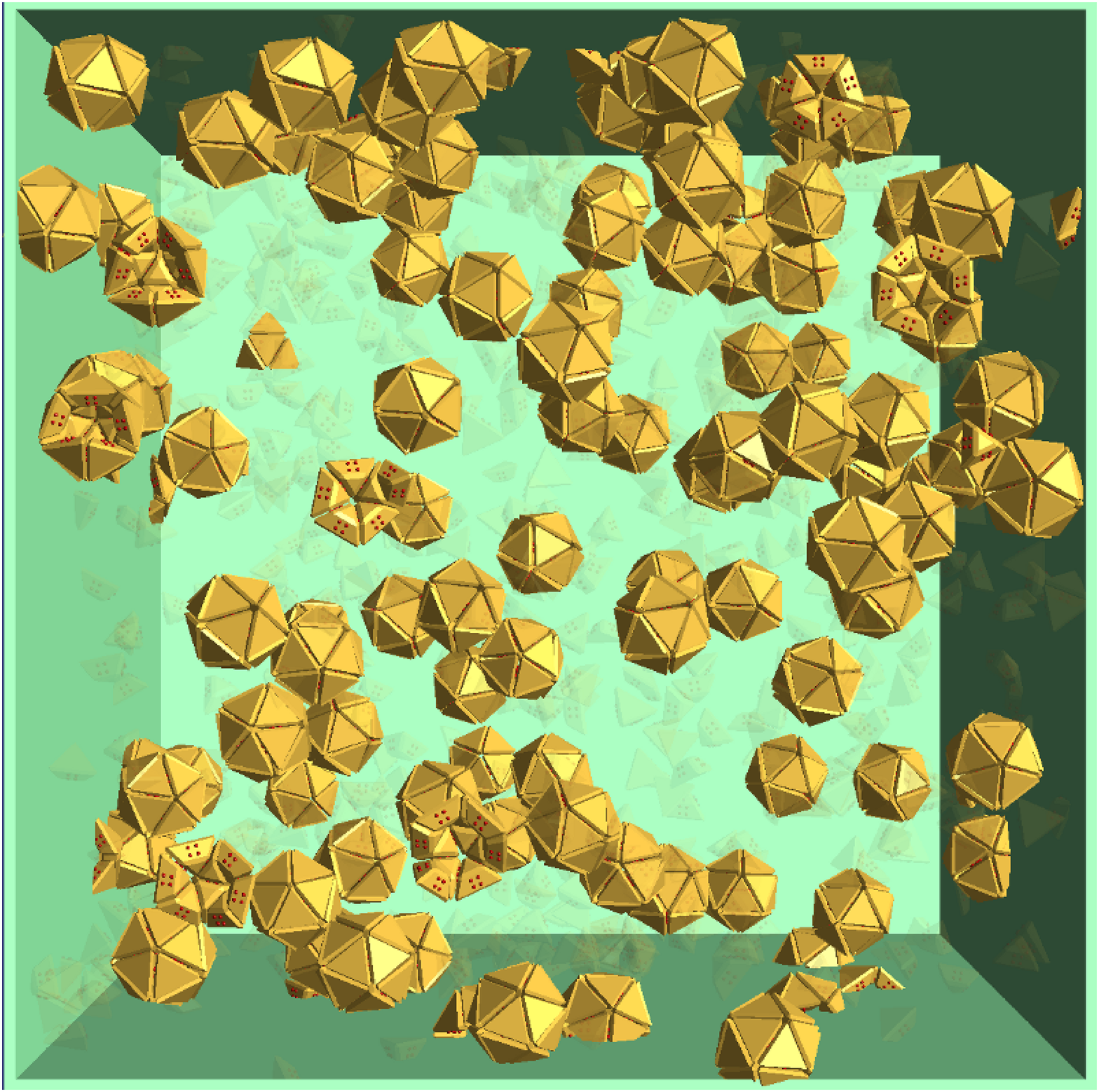}
\end{center}
\caption{\label{fig:12} (Color online) Image of a system (see text) with 105
complete shells; the remaining particles are shown semi-transparently and the
solvent is omitted (apparently open shells are visual artifacts due to periodic
boundaries).}
\end{figure}

Fig.\,\ref{fig:12} shows the outcome of a run with 2750 triangular particles,
together with solvent, at a lower density ($\rho = 0.1$) and with $e = 0.14$;
after $45 \times 10^6$ steps 105 out of a possible 137 closed shells have
formed. Other partial structures (and monomers) are shown semi-transparently.
Note that periodic boundaries are applied at the level of individual particles,
so that shells crossing the region boundaries appear fragmented; solvent
particles are not shown here (for clarity), but they fill the volume.

The stability of complete shells was tested by extending this particular run,
once it had produced the large shell population, after reducing $e$ to a value
where assembly would normally yield only a few dimers at most. The residual
small clusters promptly vanished, followed by the gradual disappearance of
larger partial shells; eventually only the original fully closed shells
remained, together with the occasional dimer. This is a demonstration of
hysteresis \cite{sin03}, the enhanced stability and consequent survival of
complete shells, even when growth conditions become unfavorable.

\subsection{Growth events}

The growth histories of individual shells can be followed, but this requires
associating partial assemblies at different stages of the run with particular
shells at run's end. Since particles join and leave different growing clusters,
sometimes multiple times, the relevant cluster is the one having maximum
population overlap with the final shell. The criterion is not perfect, since a
given final shell can own the majority of particles from more than one smaller
early cluster, and the identity of the cluster containing the most particles
destined for a given shell can change; but events of this kind generally have
only minimal influence on the ability to monitor individual cluster histories
once a relatively stable core has formed, and they become even less of a concern
at more advanced stages of growth. Cluster analysis considers all the current
members, including particles that subsequently detach.

\begin{figure}
\begin{center}
\includegraphics[scale=0.90]{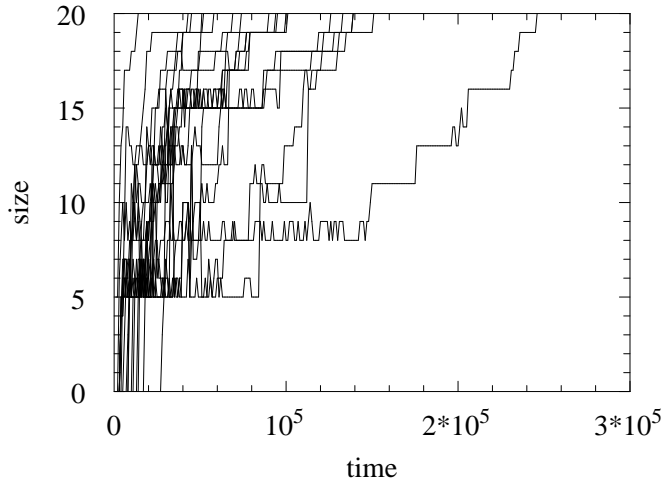}
\end{center}
\caption{\label{fig:13} Size histories for 20 of the shells.}
\end{figure}

The growth histories of a subset of 20 (of the total of 83) complete shells, for
$e=0.13$, are shown in Fig.\,\ref{fig:13}. Initial growth to pentamer size is
rapid, but subsequent growth rates exhibit a broad distribution. While there are
clusters that develop rapidly -- some even monotonically -- to completion, the
paths of others become temporarily blocked, repeatedly adding and then promptly
losing an additional particle until a more lasting growth step is eventually
taken.

\begin{figure}
\begin{center}
\includegraphics[scale=0.90]{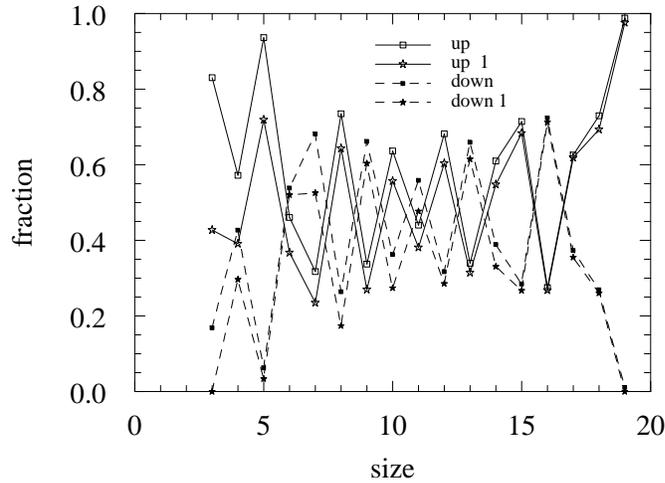}
\end{center}
\caption{\label{fig:14} Fraction of size changing (up or down) events occurring
for clusters of each size; unit size changes are also shown separately.}
\end{figure}

The statistics of the different types of size-changing events offers further
insight into the growth process. Fig.\,\ref{fig:14} shows the fractions of events
corresponding to unit size changes in each direction, together with the
fractions of all size-changing events irrespective of magnitude, each as a
function of cluster size. The reversible nature of the process is abundantly
clear. A substantial fraction of events at nearly all sizes (excluding 5 and 19)
involve size decreases. Moreover, there are cluster sizes (e.g., 7, 9, 13, 16)
for which the size is actually more likely to decrease than increase. So while
unit size changes tend to account for the majority of events, the process is
bidirectional. This is clear evidence of the strong effect of reversibility,
another key outcome of the MD simulations. The consequences of reversibility
have also been considered \cite{zlo94} using reaction kinetics. One of the
functions of reversibility is to provide a mechanism for the error correction
essential to avoiding kinetic traps. The presence of reversibility is also an
indication that self-assembly can occur in a near-equilibrium state, with only a
slight bias in favor of growth.

Cluster lifetimes can also be measured \cite{rap10}. The mean total time that
clusters exist at a given size correlates with the preferred direction of size
change, and is greater at those sizes where a size increase is more likely than
a decrease. Owing to the reversibility of bond formation, the time a cluster
spends at a particular size tends to be made up of several distinct intervals.
From measurements of the mean uninterrupted time at a given size -- a
substantially shorter period for most sizes -- it is possible to estimate the
typical number of times a cluster attains that particular size; the values were
found to range from a low of almost unity at size 19, up to about 15 at size
five. Finally, the uninterrupted times can be examined individually according to
the direction of the next size change; where there is a distinct difference, it
follows the same trend as the total time. Results of this kind are further
consequences of a growth process dominated by reversibility.

\subsection{Intermediate states}

The nature of the intermediate states along the growth pathways is particularly
interesting \cite{rap08}. Table~\ref{tab:2} summarizes a series of measurements
at $5 \times 10^5$ step intervals, for $e=0.13$. The clusters, when grouped by
bond count, reveal an extremely strong preference for maximally bonded (minimal
energy) states; all the remaining clusters are within two bonds of maximum. The
table includes the total numbers of possible clusters -- equivalent to the
distinct connected embeddings of triangles in an icosahedral lattice, quantities
that are readily computed \cite{rap87} -- almost all of which never appear among
the configurations observed. Thus, for example, while almost 92\% of clusters of
size 12 adopt the single 15 bond form, and the rest have either 14 or 13 bonds,
none of the other 446 realizations with fewer bonds are actually observed. The
effect of an imposed preference for maximally bonded intermediates has been
studied with reaction kinetics \cite{end05}; the MD simulations reveal this
preference to be intrinsic, another key result.

\begingroup
\begin{table}
\caption{\label{tab:2} Intermediate states along the growth pathways (from
\cite{rap08}): the mean cluster fractions ($f$) are grouped by size ($s$) and
bonds ($b$), with the maximally bonded fractions shown in bold; the numbers of
distinct cluster realizations ($n$) are included for comparison, and the final
columns enumerate the possible realizations that were not observed (sizes with
unique bond counts are omitted).}
\begin{center}
\begin{tabular}{r@{\ \ \ \ \ }rrcrrcrrc@{\ \ \ \ \ \ \ \ }rr}
\hline
      & \multicolumn{9}{c}{Observed} &         \multicolumn{2}{c}{Others}  \\
  $s$ &$b\ $ &$n$& $f$  & $b\ $ &$n$ & $f$ &$b\ $& $n$ & $f$   & $b\ $  & $n$ \\
\hline
   5  &  5: & 1 & {\bf 0.948} &  4: &  5 & 0.052 &     &     &       &        &     \\
   6  &  6: & 1 & {\bf 0.953} &  5: & 13 & 0.047 &     &     &       &        &     \\
   7  &  7: & 4 & {\bf 0.979} &  6: & 22 & 0.021 &     &     &       &        &     \\
   8  &  9: & 1 & {\bf 0.851} &  8: & 11 & 0.140 &  7: &  46 & 0.009 &        &     \\
   9  & 10: & 3 & {\bf 0.938} &  9: & 27 & 0.062 &     &     &       &     8: &  79 \\
  10  & 12: & 1 & {\bf 0.808} & 11: & 13 & 0.166 & 10: &  60 & 0.026 &     9: & 151 \\
  11  & 13: & 3 & {\bf 0.931} & 12: & 28 & 0.069 &     &     &       & 11-10: & 328 \\
  12  & 15: & 1 & {\bf 0.917} & 14: & 11 & 0.073 & 13: &  74 & 0.010 & 12-11: & 446 \\
  13  & 16: & 4 & {\bf 0.876} & 15: & 31 & 0.105 & 14: & 142 & 0.019 & 13-12: & 372 \\
  14  & 18: & 1 & {\bf 0.802} & 17: & 15 & 0.198 &     &     &       & 16-13: & 417 \\
  15  & 20: & 1 & {\bf 0.825} & 19: &  5 & 0.146 & 18: &  38 & 0.029 & 17-15: & 170 \\
  16  & 21: & 4 & {\bf 0.915} & 20: & 19 & 0.068 & 19: &  38 & 0.017 &    18: &  28 \\
  17  & 23: & 1 & {\bf 0.923} & 22: &  7 & 0.077 &     &     &       &    21: &  12 \\
  18  & 25: & 1 & {\bf 0.888} & 24: &  5 & 0.112 &     &     &       &        &     \\
\hline
\end{tabular}
\end{center}
\end{table}
\endgroup

\subsection{Visualizing shell assembly}

\begin{figure*}
\begin{center}
\includegraphics[scale=0.60]{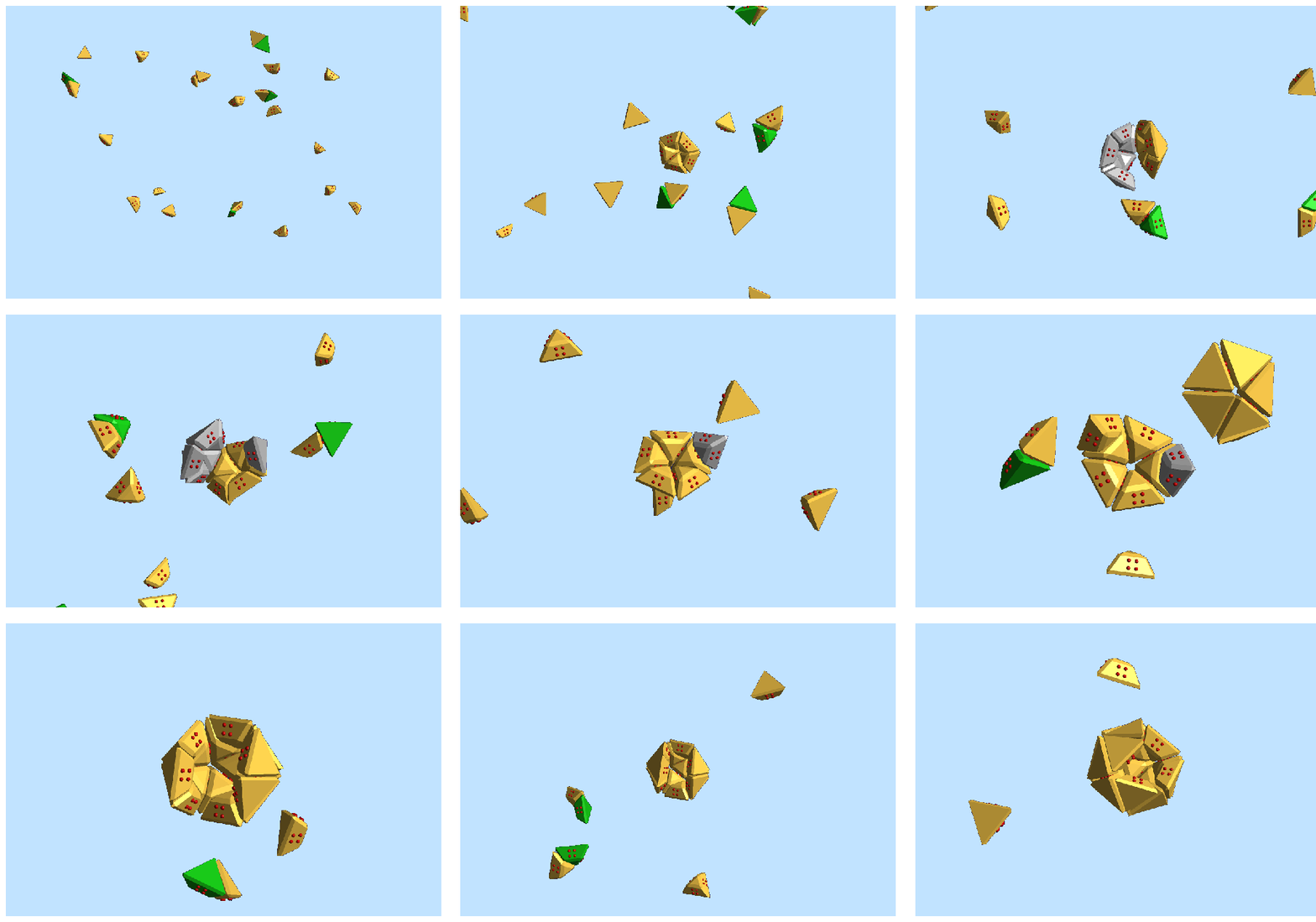}
\end{center}
\caption{\label{fig:15} (Color online) Stages in the reversible growth of one of
the shells, including two pentamers being temporarily bonded (upper-right
frame); only the particles directly involved are included although some lie
outside the field of view (solvent is not shown); the color coding is explained
in the text.}
\end{figure*}

Imagery is especially helpful for detailed exploration of those aspects of the
growth process that are less readily quantifiable, and may suggest additional
approaches for analyzing the pathway details. Fig.\,\ref{fig:15} shows a series
of frames covering several stages in the growth of just one of the shells. Only
the particles directly involved are included (although some may be too far away
to appear in the frames). Suitable color coding, based on the known final shell
membership specifies the eventual disposition of the particles: yellow for
particles destined for (or already in) the final shell, gray for particles only
temporarily attached to the growing shell, and green for particles that are
temporarily attached to yellow particles not yet in the final shell.

One such growth sequence was discussed in Ref.\,\cite{rap10}. The sequence shown
here exhibits very different behavior, with two pentamer-size clusters bonded,
and after a certain period of time breaking apart, with only one destined for
the final shell. Subsequently, it links up with a different pentamer and then
continues to grow by smaller increments. The last frame of the sequence shows
the final particle about to close the shell. Shells clearly have very distinct
histories. Particles, literally, can come and go; only when bound to most of its
neighbors, and embedded in a substantial portion of a partial shell, is an
individual particle unlikely to be knocked out of position. Population exchanges
of this kind are not readily characterized in a quantitative manner, but can be
tracked visually.

\subsection{Assembly of larger shells}

The next stage in developing the approach is modeling the self-assembly of
larger shells in a solvated environment. Current efforts are focused on
pentakisdodecahedral and T=1 shells, built from 60 triangular or trapezoidal
particles, respectively. The systems contain 2750 particles, sufficient in
principle for 45 shells, with other aspects of the simulations, including the
interaction parameters and overall system size, remaining unchanged. While
growing such shells without a solvent -- and therefore subject to artificial
breakup and damping issues as described above -- was not especially demanding
computationally, the extra effort resulting from inclusion of the solvent is
substantial. The simulation runs will have to be considerably longer than for
the icosahedra in order to achieve respectable yields, and since the shell size
is tripled, the number of particles in the system, as well as the number of
solvent atoms, will eventually need to be increased accordingly, although this
is not reflected in the present exploratory results.

\begin{figure*}
\begin{center}
\includegraphics[scale=0.40]{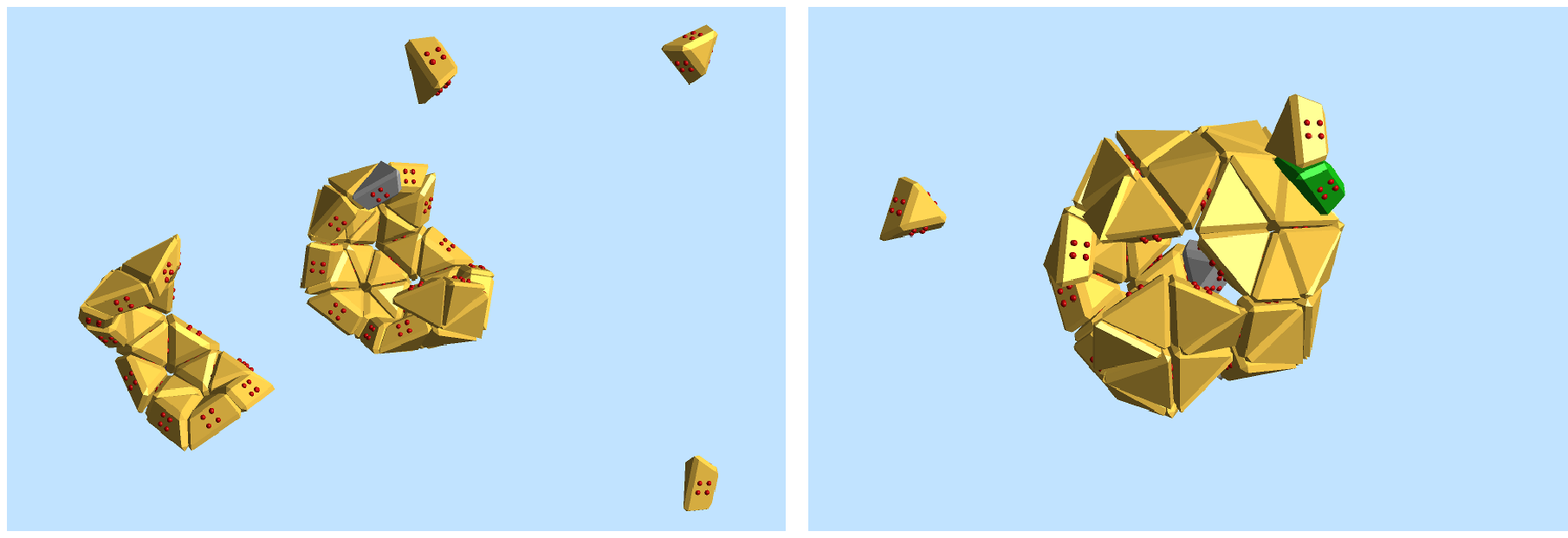}
\end{center}
\caption{\label{fig:16} (Color online) Stages in the merging of two
subassemblies after 20, 21 and $22 \times 10^6$ steps, and the 54-member, still
incomplete shell after $33 \times 10^6$ steps (the color coding is the same as
before).}
\end{figure*}

The first of two examples of what occurs along the growth pathway is shown in
Fig.\,\ref{fig:16}, where two substantial subassemblies bond together. The
opportunity for complex events of this kind is increased for larger shells, with
reversible bonding helping to ensure -- although not guaranteeing -- that
obstructing particles are ejected from the cluster.

\begin{figure*}
\begin{center}
\includegraphics[scale=0.54]{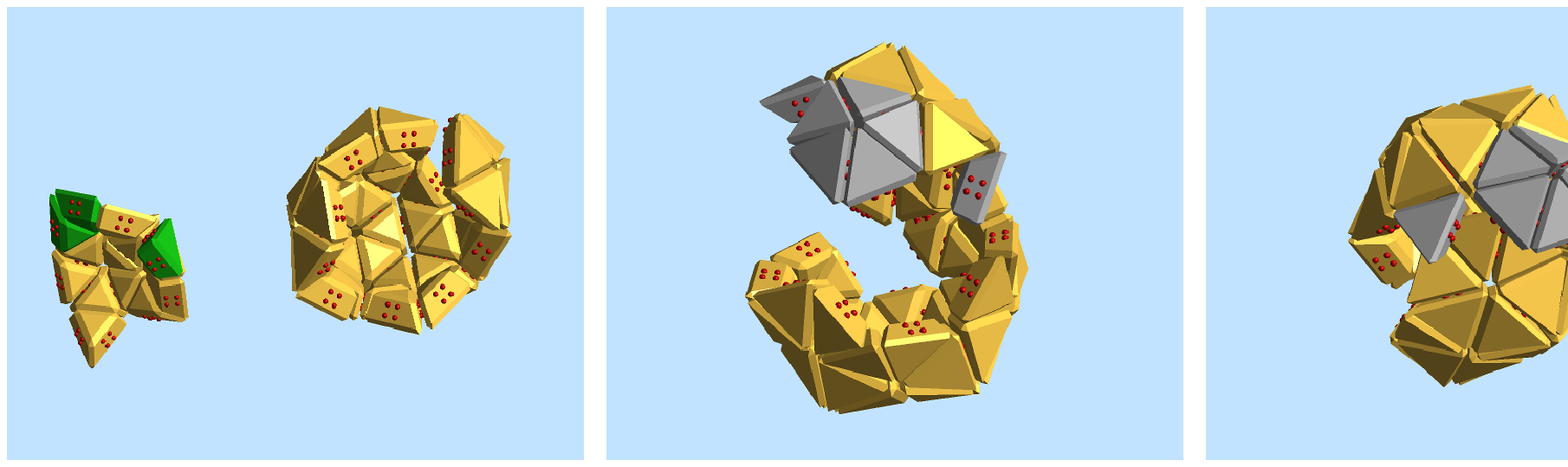}
\end{center}
\caption{\label{fig:17} (Color online) Temporary formation of an incorrect
structure: at $29.5 \times 10^6$ steps the main subassembly (size 41) is about
to merge with a smaller piece; at $29.9 \times 10^6$ steps the merge has
occurred (size 59); the cluster at $30.6 \times 10^6$ steps (size 58) with
overlapping shell components clearly visible; at $34.3 \times 10^6$ steps the
extra pieces have broken off leaving a cluster of size 54.}
\end{figure*}

A merging event followed by serious error correction, involving a different
cluster, appears in Fig.\,\ref{fig:17}. As a result of the merge, the growing
shell acquires an unwanted pentamer that is seen to inhibit future successful
growth; it remains attached for a considerable period of time ($5 \times 10^6$
steps) before breaking off and allowing the shell to continue along its growth
path. The message conveyed by these two sequences reinforces the earlier
evidence of cluster histories being subject to wide variation.

\begin{figure}
\begin{center}
\includegraphics[scale=0.20]{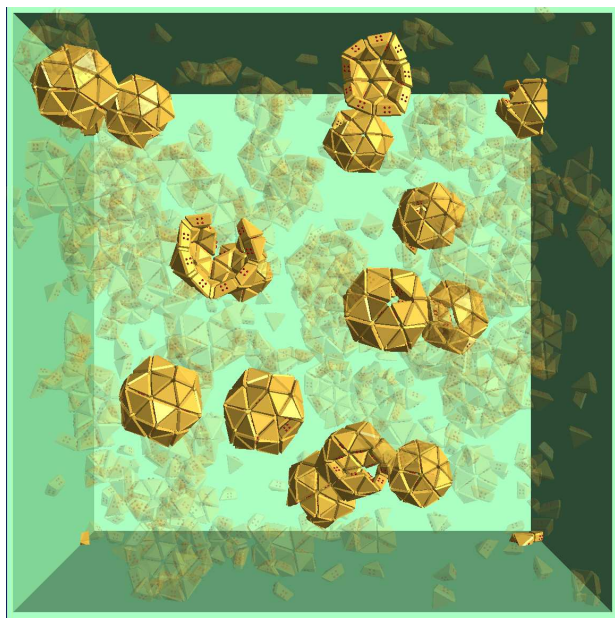}
\end{center}
\caption{\label{fig:18} (Color online) Shells growing in solvent, including two
that are complete (size 60); clusters with size $< 50$ and monomers are shown
semi-transparently and the solvent is omitted.}
\end{figure}

The final image, Fig.\,\ref{fig:18} shows the state of a system of triangular
particles (at the time of writing) after $44 \times 10^6$ steps. There are 13
clusters of size $\ge 50$, including two closed shells, the first of which
completed after $39 \times 10^6$ steps, and one of size 59 (none of these three
correspond to the structure examined in Fig.\,\ref{fig:17}). At this stage 200
monomers remain, so further shell completion appears likely. The largest shell
size in an analogous simulation with trapezoidal particles was 47, after $19
\times 10^6$ steps, but the monomer supply was practically exhausted. The
completion rate here is, not surprisingly, much slower than for the smaller
shells which, after runs of this duration, had already yielded significant
numbers of complete structures.

\section{Conclusion}

The present simulations provide a demonstration that a simple potential energy
function, based on structural considerations, is essentially all that is
required to drive self-assembly. This
approach, whose focus is on examining the essential physics of the growth
process,
allows the influence of shape and
interactions to be examined relatively easily, in contrast to the heavy
computational demands required for a more detailed atomic description.

Self-assembly at submicroscopic scales, where intrinsic thermal fluctuations
become important, is seen to be very different from inherently unidirectional
macroscopic assembly. Reversibility makes itself felt along the entire assembly
pathway, with dissociation more likely than growth throughout much of the
process. The surprising aspects of the results -- given the absence of any {\em
a priori} theoretical expectations -- are the fast growth rates, high yields,
and the avoidance of incorrect final structures.

Although the models are not representative of real capsomer proteins, and no
attempt has been made to develop a more explicit relationship, if there exist
universal features underlying capsid assembly, simplified systems of this kind
ought to embody their essence. It goes without saying that the more general
aspects of self-assembly suggested by these simulations, in particular the
coexistence of reversibility with a high error-free yield, are likely to have
important implications for understanding supramolecular assembly in general, and
capsid formation in particular.

\section*{References}

\bibliography{capsmodel}

\bibliographystyle{unsrt}

\end{document}